# A Machine Learning-based Non-precipitating Clouds Estimation for THz Dual-Frequency Radar

Kazuhiko Tamesue
Faculty of Science and Engineering
Waseda University
Tokyo, Japan
ktamesue@aoni.waseda.jp

Zheng Wen
Waseda University
Faculty of Science and Engineering
Tokyo, Japan
robinwen@fuji.waseda.jp

Shotaro Yamaguchi
Graduate School of Fundamental Science and Engineering
Waseda University
Tokyo, Japan
yama94sho@ruri.waseda.jp

Hiroyuki Kasai
Faculty of Science and Engineering
Waseda University
Tokyo, Japan
hiroyuki.kasai@waseda.jp

Wataru Kameyama
Faculty of Science and Engineering
Waseda University
Tokyo, Japan
wataru@waseda.jp

Toshio Sato
Faculty of Science and Engineering
Waseda University
Tokyo, Japan
toshio4.sato@aoni.waseda.jp

Yutaka Katsuyama
Faculty of Science and Engineering
Waseda University
Tokyo, Japan
katsuyama@aoni.waseda.jp

Takuro Sato
Faculty of Science and Engineering
Waseda University
Tokyo, Japan
t-sato@waseda.jp

Takeshi Maesaka
National Research Institute for Earth Science and Disaster Resilience (NIED)
Ibaraki, JAPAN
maesaka@bosai.go.jp

***Abstract*— Accurate measurement of non-precipitable clouds is important for early prediction of heavy rainfall disasters caused by extreme weather events. However, microwave cloud radar cannot observe the early stages of cloud development from non-precipitation clouds (cumulus) to cumulonimbus. In this paper, we propose a terahertz dual-frequency cloud radar using 150 GHz and 95 GHz bands to detect cloud particles in cumulus smaller than 10 μm. Using a dataset generated by the ITU-R radio propagation model, we estimate the liquid water content of non-precipitation clouds and water vapor content in atmospheric gases, respectively, by using a machine learning-based approach. The effectiveness of using the dual wavelength ratio as an explanatory variable is examined.**

## I. Introduction

Recent climate change studies predict that the frequency of heavy rainfall (or the ratio of heavy precipitation to total precipitation) is likely to increase with global warming. In predicting heavy rainfall cumulonimbus clouds, it is important to know the liquid water content (LWC) prior to cumulonimbus cloud development and the distribution of non-precipitation clouds (e.g., cumulus), which are the source of cumulonimbus clouds [1].

Conventional cloud radars using microwaves such as S-band and X-band are used to observe precipitation distribution. However, due to their long wavelengths, they are unable to capture minute cloud particles of less than 10 μm in cumulus before they develop into cumulonimbus clouds. Dual-frequency radar using Ka-band (35 GHz), W-band (94 GHz), and their dual wavelength ratio (DWR) has been reported to produce more accurate LWC detection [2, 3].

This paper reports an approach and results of a machine learning (ML) method for estimating the liquid water content (LWC) and the water vapor content (WVC) in atmospheric gases by using the radar reflectivities at two new frequencies, W-band (95 GHz) and G-band (150 GHz), and also their DWR as features in the proposed dual-frequency terahertz (THz) radar system. The approach and results of the estimation of LWC and WVC by machine learning (ML) are presented.

Recent study on modeling cloud-related processes has reported some examples of the application of machine learning (ML) to cloud detection and classification [4]. There are two main advantages in implementing machine learning. The first is the separation of attenuation due to the atmosphere from attenuation due to clouds and precipitation. The second is estimation of water vapor content between radar and clouds. Since DWR includes both the effects of attenuation due to the atmosphere and the attenuation due to clouds and precipitation, it is not possible to analytically separate the effects due to the atmosphere from those due to clouds and precipitation from the observed DWR. In addition, since there is no scattering of radio waves from the cloudless atmosphere, the DWR for the cloudless region cannot be obtained. However, the observed DWR does include the effects of atmospheric attenuation between the radar and clouds. Because atmospheric water vapor is not randomly distributed; to some extent, its distribution depends on altitude. Under these assumptions, machine learning estimation is effective.

This paper is organized as follows: Section II describes the methodology for THz dual-frequency radar system design and path attenuation modeling, Section III describes the generation of training data for the ML-based path model and the evaluation results of the simulations. Note that the effects of rain on THz radar are not included in this paper since rain attenuation is significant for THz radar, and finally we conclude in Section IV.

## II. Methology

### A. System Design

Fig. 1 shows an example of the pass-through attenuation vs. frequency response of atmospheric gases in summer in the mid-latitudes as defined in ITU-R P.835-6 [5]. The pass-through attenuation characteristics in Fig. 1 are based on the calculation algorithm in ITU-R P.676-13 [6]. Below 300 GHz, oxygen absorption is present at 60 and 120 GHz, and water vapor absorption at 22 and 180 GHz. In between, 95 GHz and 150 GHz are favorable frequency bands for DWR sensitivity. The diameter of cloud particles in non-precipitating clouds ranges from 1 μm to 10 μm, and the LWC where cloud particles grow into raindrops is 0.5 g/m$^3$ to 1.0 g/m$^3$. 150 GHz band is suitable for treating small raindrops in non-precipitating clouds as Rayleigh scattering. In previous studies, 94 GHz cloud radars of W-band have been used in meteorological installations such as W-band ARM Cloud Radar (WACR) to detect atmospheric profiles, cloud base altitude, and LWC with high accuracy [2]. On the other hand, there are few studies of G-band cloud radar except for a study of differential absorption radar for water vapor profiling at 170 GHz has been reported [7].

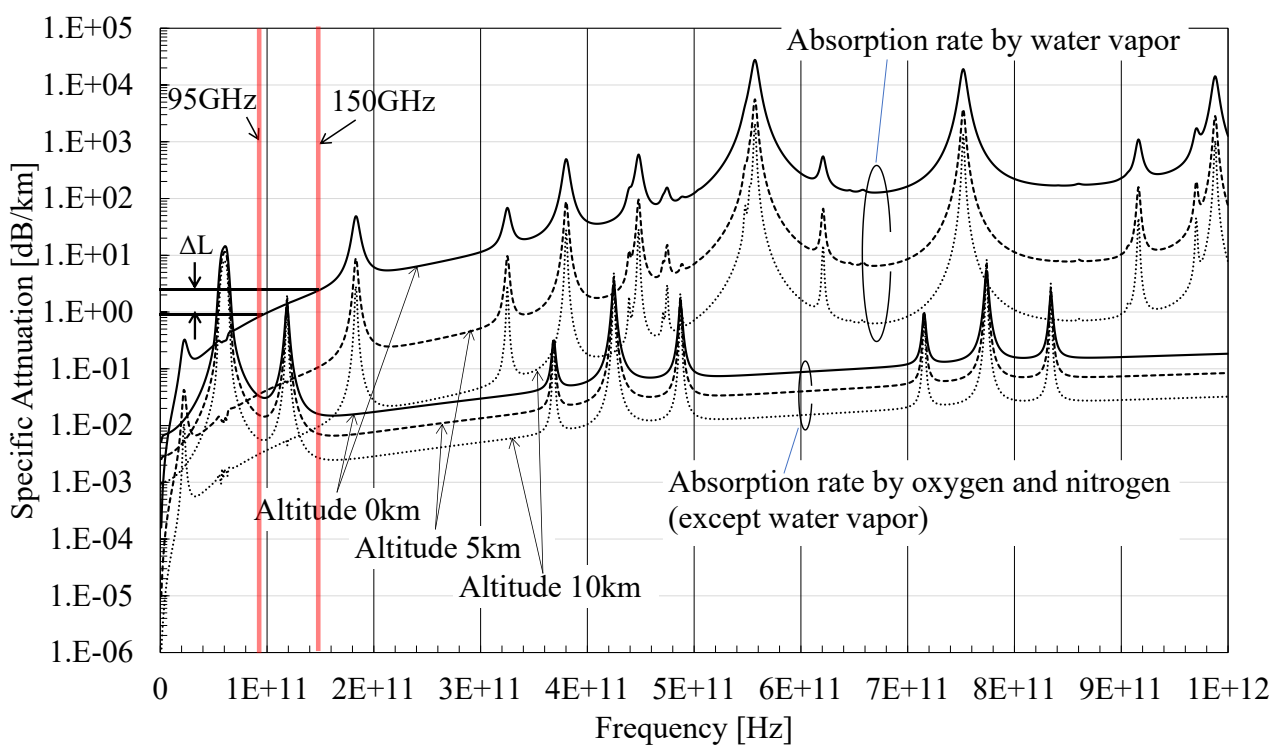

Fig. 1. Attenuation Rate vs. Frequency of Atmospheric Gases (mid-latitude, summer, attenuation by water vapor only, attenuation by oxygen other than water vapor, etc.)

Fig. 2 shows an example of cloud transit attenuation vs. frequency in summer over a mid-latitude region defined by ITU-R P.840-8 [8]. Relative humidity is set to 50% and 25% at altitudes of 1-2 km and 4-5 km, respectively. The WVC is calculated from relative humidity by finding the saturated water vapor density using Tetens' approximation from the pressure and temperature at each altitude.

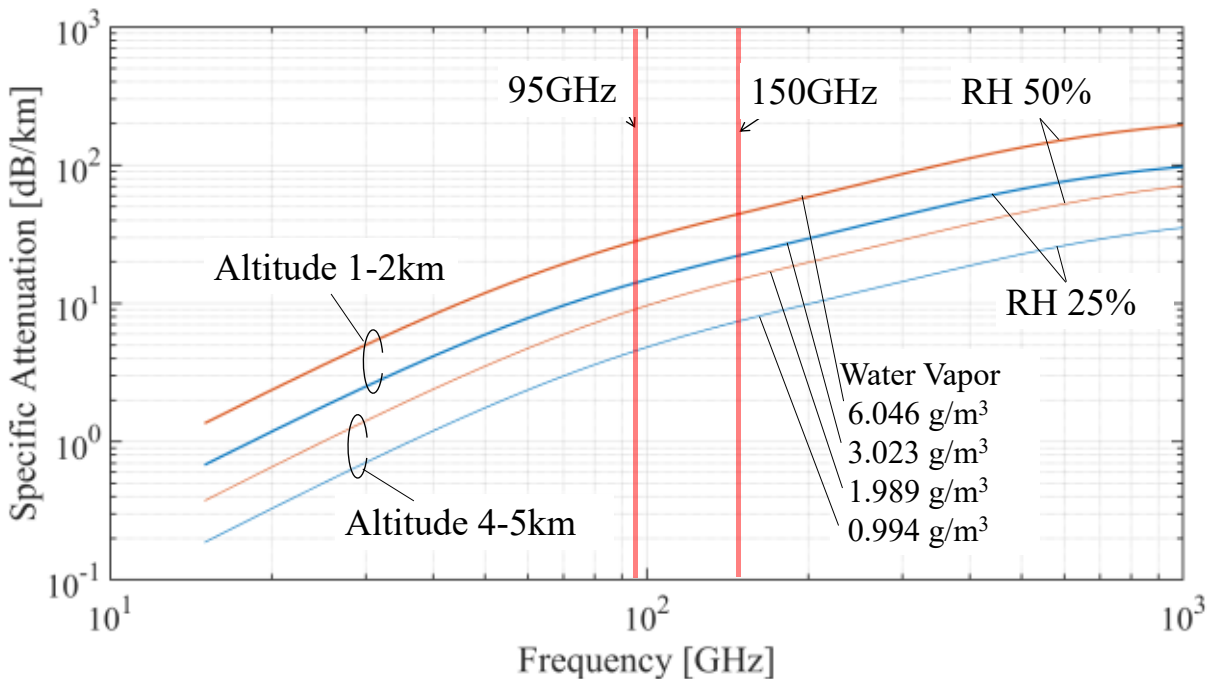

Fig. 2. Cloud Transit Attenuation vs. Frequency (mid-latitude, summer, cumulus altitude 1-2 km, 4-5 km, relative humidity 50% and 25%)

The radar reflectivity Z of the cloud radar is given by (1) where D is the diameter of the cloud particle. Assuming Rayleigh scattering, the radar scattering cross section σ of a cloud particle is given by (2) where λ is the wavelength and K is a constant. The total of the scattering cross sections of cloud particles per unit volume, η, is given by (3). Since the received radar power is proportional to η, it is also proportional to the radar reflectivity Z.

$$Z = \sum D^6 \tag{1}$$

$$\sigma = \frac{\pi^5}{\lambda^4}(K^2 D^6) \tag{2}$$

$$\eta = \sum \sigma = \frac{\pi^5}{\lambda^4} K^2 \sum D^6 = \frac{\pi^5}{\lambda^4} K^2 Z \tag{3}$$

Fig. 3 shows a schematic diagram of the dual-frequency THz radar proposed in this paper.

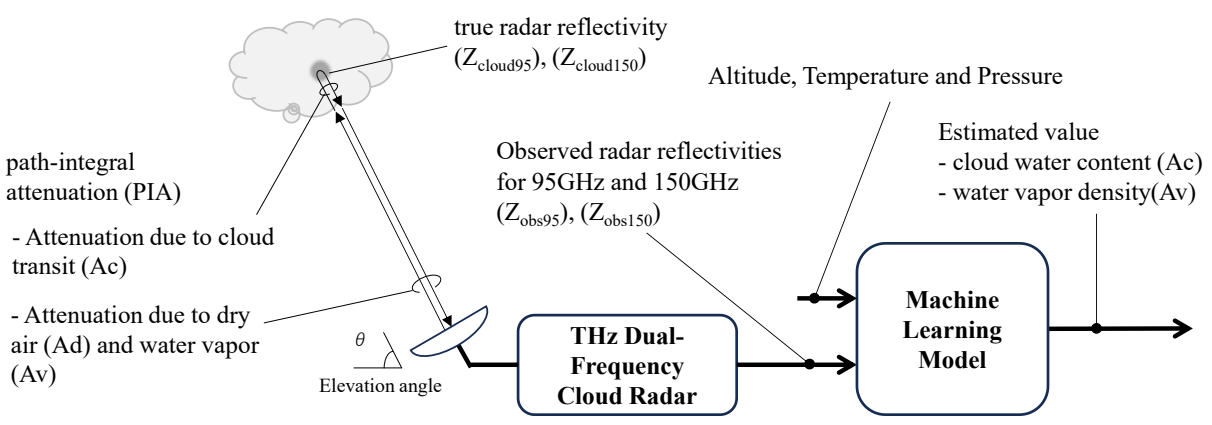

Fig. 3. Conceptual Block Diagram of THz Dual-frequency Radar

Fig. 4 shows the characteristics of cloud reflectivity of vs. LWC [9]. Rainfall is formed when LWC exceeds 1 g/m$^3$. The radar reflectivity of the cloud at the beginning of that rainfall is -13 dBZ. To observe cumulus, a sensitivity of the radar is required to be higher than that value. The radar to be developed uses two frequencies, 95 GHz and 150 GHz.

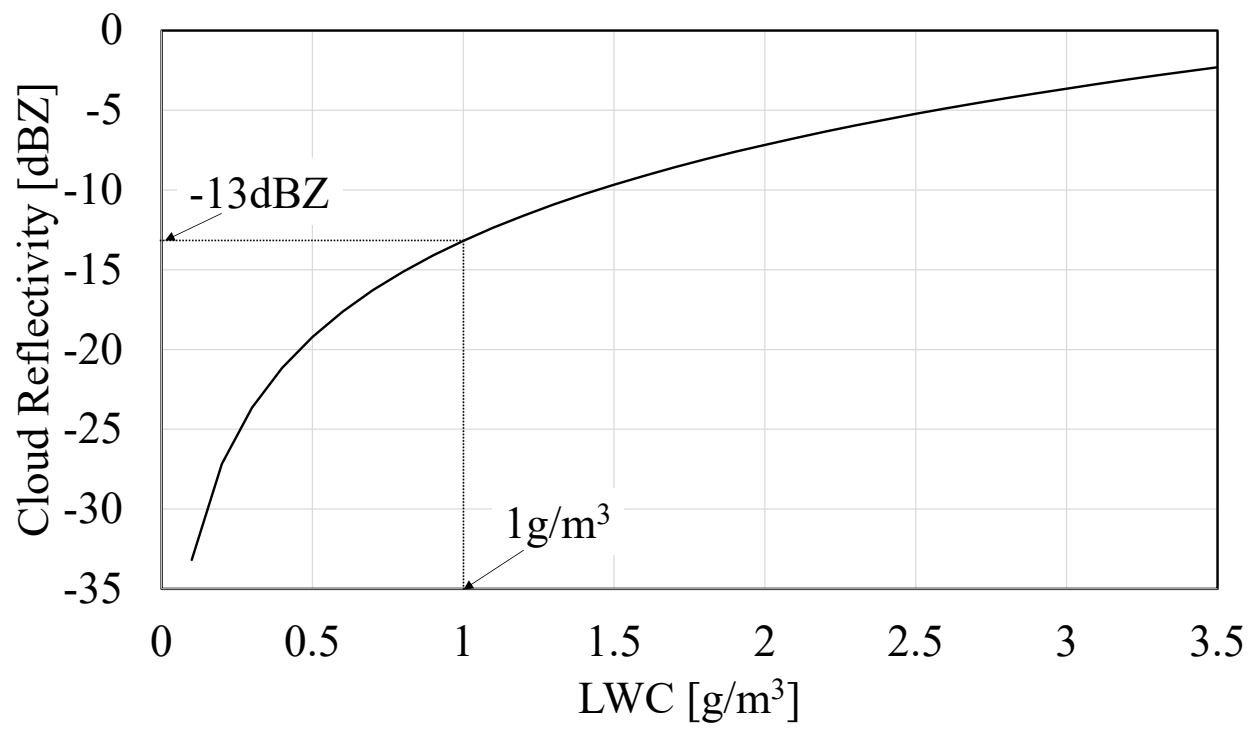

Fig. 4. Cloud Reflectivity vs. LWC

## *B. Path Gain Model*

### *1) Radar reflectivity and path attenuation*

Assuming a scenario where the elevation angle θ is 90°, radio waves are emitted in the vertical direction and radar reflectivity from clouds is observed.

The true radar reflectivities Zcloud95 and Zcloud150 cannot be observed directly because the observed radar reflectivities Zobs95 and Zobs150 must take into account the effects of the observed target's own clouds and atmospheric gases transit attenuation. It is calculated from (4) by adding the path-integral attenuation (PIA) to the observed radar reflectivity Zobs:

$$\begin{cases} PIA(h,d) = 2\int_0^h \{A_c(h,d) + A_d(h) + A_v(h)\}dh \\ Z_{cloud} = Z_{obs} + PIA(h,d) \end{cases} \tag{4}$$

where d is cloud thickness, and h is cloud altitude; and the attenuation due to cloud transit, dry air, and water vapor are denoted by Ac(h, d), Ad(h), and Av(h), respectively. The PIA indicates the integral from ground level to altitude h.

The transit attenuation Ad(h) and Av(h) due to dry air and water vapor depend on the temperature T, atmospheric pressure P, and water vapor density at altitude h.

*2) Discussion of path attenuation*

The attenuation rate of atmospheric gases per unit distance is expressed as the sum of the attenuation rates due to water vapor and oxygen, etc. as defined in ITU-R P.676-13 [6], shown in Fig. 2. The reason why each curve shown in Fig. 2 shows different atmospheric attenuation rates at different altitudes is because of the different altitude distributions of temperature, pressure, and water vapor density.

For the reference standard atmosphere defined in ITU-R P.835-6 [5], the temperature, pressure, and relative humidity are shown in (5) to (7) between the ground and altitude h = 11 km where h' is the geopotential altitude. The reference standard atmosphere defined in ITU-R P.835-6 [5] is the monthly mean of the vertical distributions of temperature, pressure, and relative humidity at 353 locations around the world for 10 years. The reference standard atmosphere is an approximation calculated from radiosonde observations of temperature, pressure, and relative humidity at 353 sites around the world over a 10-year period (1980-1989).

$$h = \frac{6356.766h'}{6356.766-h'} \tag{5}$$

$$T(h') = 288.15 - 6.5h' \quad for\ 0 \leq h' \leq 11 \tag{6}$$

$$\begin{cases} \rho(h) = \rho_0 \exp(-h/h_0) & g/m^3 \\ \rho_0 = 7.5 & g/m^3 \\ e(h) = \rho(h)T(h)/216.7 & hPa \end{cases} \tag{7}$$

From (5) to (7), temperature T, pressure e, and water vapor density ρ are closely correlated with altitude h. Therefore, WVC and LWC can be estimated by retrieval from the attenuation Av(h) due to atmospheric gases and the attenuation Ac(h, d) through clouds, respectively.

On (6) and (7), since altitude, temperature, pressure, and water vapor density vary at each observation location, it is preferable to generate them from a function fitted based on radiosonde and other observation data for each region. For example, the geopotential altitude of 11 km, meaning the boundary between the troposphere and stratosphere, varies with latitude. In low-latitude regions near the equator, the boundary is lower. In higher latitudes, it is higher. Surface water vapor density is generally higher in regions near the ocean and lower inland.

## III. MACHENE LEARNING-BASED ESTIMATION OF CLOUD WATER CONTENT

### *A. Simulation Model*

ML employs the radar reflectivities Zobs95 and Zobs150 and the altitude distribution of (temperature, pressure and water vapor density) for the explanatory variables. Attenuation Av(h) due to cloud attenuation Ac(h, d) and atmospheric gases is estimated.

The altitude is obtained from the linear distance to the cloud and the elevation angle by the radar reflectivities Zobs95 and Zobs150. Attenuation Av(h) due to atmospheric gases is a function of altitude from ITU-R P.835-6 [5]. Thus, the temperature, pressure, and atmospheric water vapor density, or the altitude at which they are correlated, are the explanatory variables. For cloud transit attenuation Ac(h, d), in addition to altitude being an explanatory variable, cloud thickness is also an explanatory variable.

We apply an ML model suitable for multiple regression analysis, in which multiple objective variables are computed from multiple explanatory variables. In this paper, we employ eXtreme Gradient Boosting (XGBoost), which is an ensemble learning method with decision trees and boosting [10]. XGBoost is suitable for predicting and explaining continuous numerical data.

### *B. Dataset and Condition*

We use the models from ITU-R P.676-13 [6] and P.840-8 [8] for the generation of atmospheric gases attenuation and cloud transit attenuation in the PIA(h, d) calculations, respectively. To obtain altitude profiles of temperature, pressure, and water vapor density, we employ the model on ITU-R P.835-6 [5]. We set the ground and one cloud and set the LWC peak at an altitude half the thickness of the cloud.

Table I lists the detailed dataset parameters and the model, where 10-fold cross validation is applied with 90% for training and 10% for evaluation. A total of 314,370 data sets are generated for various weather conditions.

All cases include altitude, temperature, pressure for explanatory variables, and LWC and WVC for objective variables. Each case includes Zobs95, Zobs150 or difference of Zobs95 and Zobs150 for explanatory variables.

TABLE I. PARAMETERS FOR DATASET AND MODEL

|  |  |
| --- | --- |
| Atmospheric pressure [hPa], temperature [°C], and water vapor [g/m$^3$] from ground up to the cloud. | - Latitude Model: Standard / Low / Mid / High<br>- Water Vapor Density (WVC) : From 14.5 to 17.5 in 1.0 step<br>- Season: Summer |
| Cloud center altitude (hc) [m]. | From 1 000 to 3800 in 200 steps |
| Cloud thickness (dc) [m]. | If 2hc-1200 < 3 000: 400 to hc in 200 steps<br>If 2hc-1 200 ≥ 3 000: 400 to 2 800 in 200 steps |
| Peak LWC in cloud [g/m$^3$]. | For hc < 2 000: 0.2 to 1.0 in 0.2 steps<br>For hc ≥ 2 000: 0.4 to 1.0 in 0.2 steps |
| ML Model | XGBoost<br>(Python Package Learning API, where parameters are default values) |

### *C. Results and Discussion*

The results of the estimation of LWC and WVC are shown from Fig. 5 to Fig. 10.

The left and right panels of Fig. 5 show the average of root mean square error (RMSE) and the average of coefficient of determination ($R^2$), respectively. The results including the three explanatory variables Zobs95, Zobs150, and Zobs95-Zobs150 show the best with the RMSE being the smallest and the $R^2$ being the largest.

Fig. 6 shows the feature importance in the fold 1, which is a measure of the importance of the explanatory variables in the estimation of LWC. The results show that the radar reflectivities Zobs95 followed by Zobs95-Zobs150, which indicates DWG, have the largest contribution.

The left and right panels of Fig. 7 show the average of RMSE and $R^2$ for the estimation of WVC in atmospheric gases, respectively. The results including the three explanatory variables $Zobs_{95}$, $Zobs_{150}$, and $Zobs_{95}$-$Zobs_{150}$ show the best results, with the lowest RMSE and the largest $R^2$.

Fig. 8 shows the feature importance of the explanatory variables in the estimation of atmospheric WVC in the fold 1. The results show that atmospheric pressure P and temperature T contribute most significantly to the results. The radar reflectivity is relatively small because scattered reflected waves do not occur at cloudless altitudes.

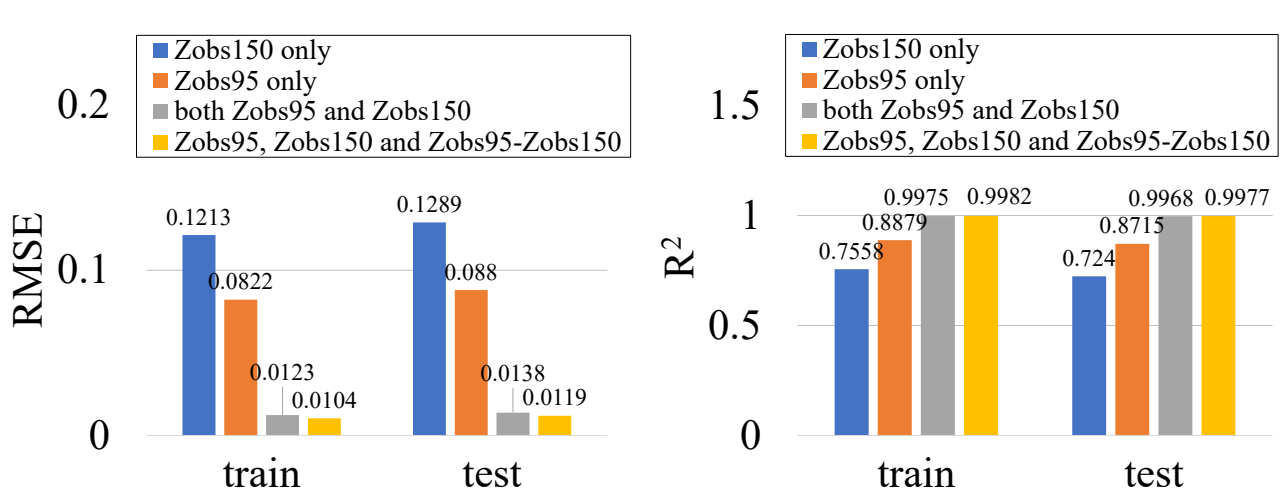

Fig. 5. Estimation Accuracy by RMSE and $R^2$ (LWC)

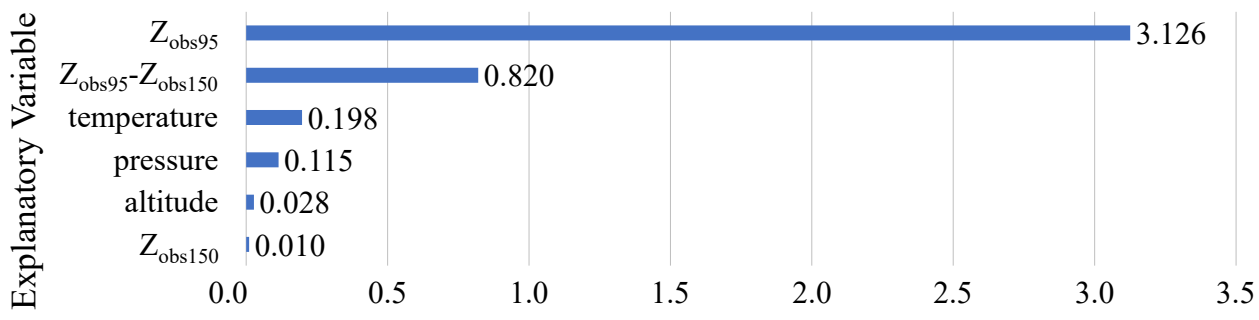

Fig. 6. Feature Importance of Each Explanatory Variable in Fold 1 (LWC)

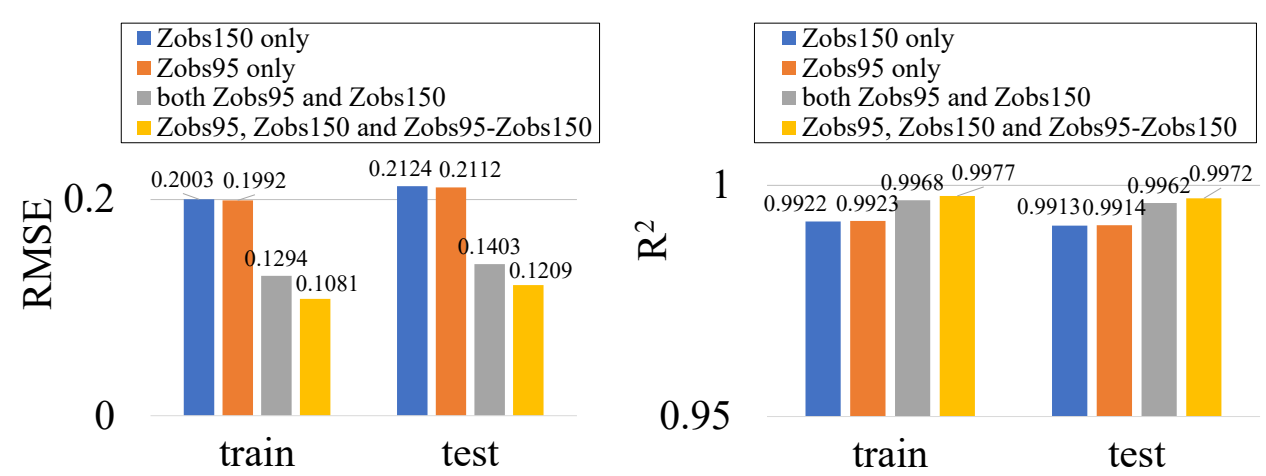

Fig. 7. Estimation Accuracy by RMSE and $R^2$ (WVC)

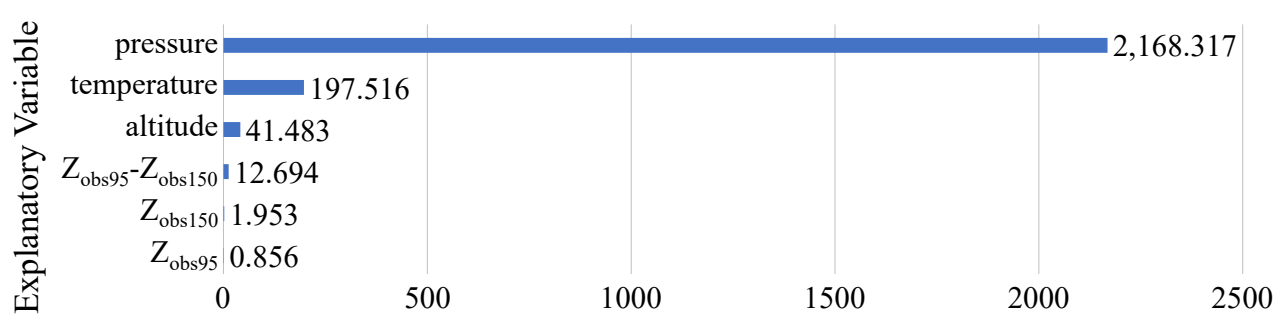

Fig. 8. Feature Importance of Each Explanatory Variable in Fold 1 (WVC)

Fig. 9 and Fig. 10 show the estimation accuracy of LWC and WVC in the fold 10. The x-axis shows the index of the evaluation results splitted by the 10-fold cross validation. The results using DWG with Zobs95 and Zobs150 show the best fitting with the actual data series.

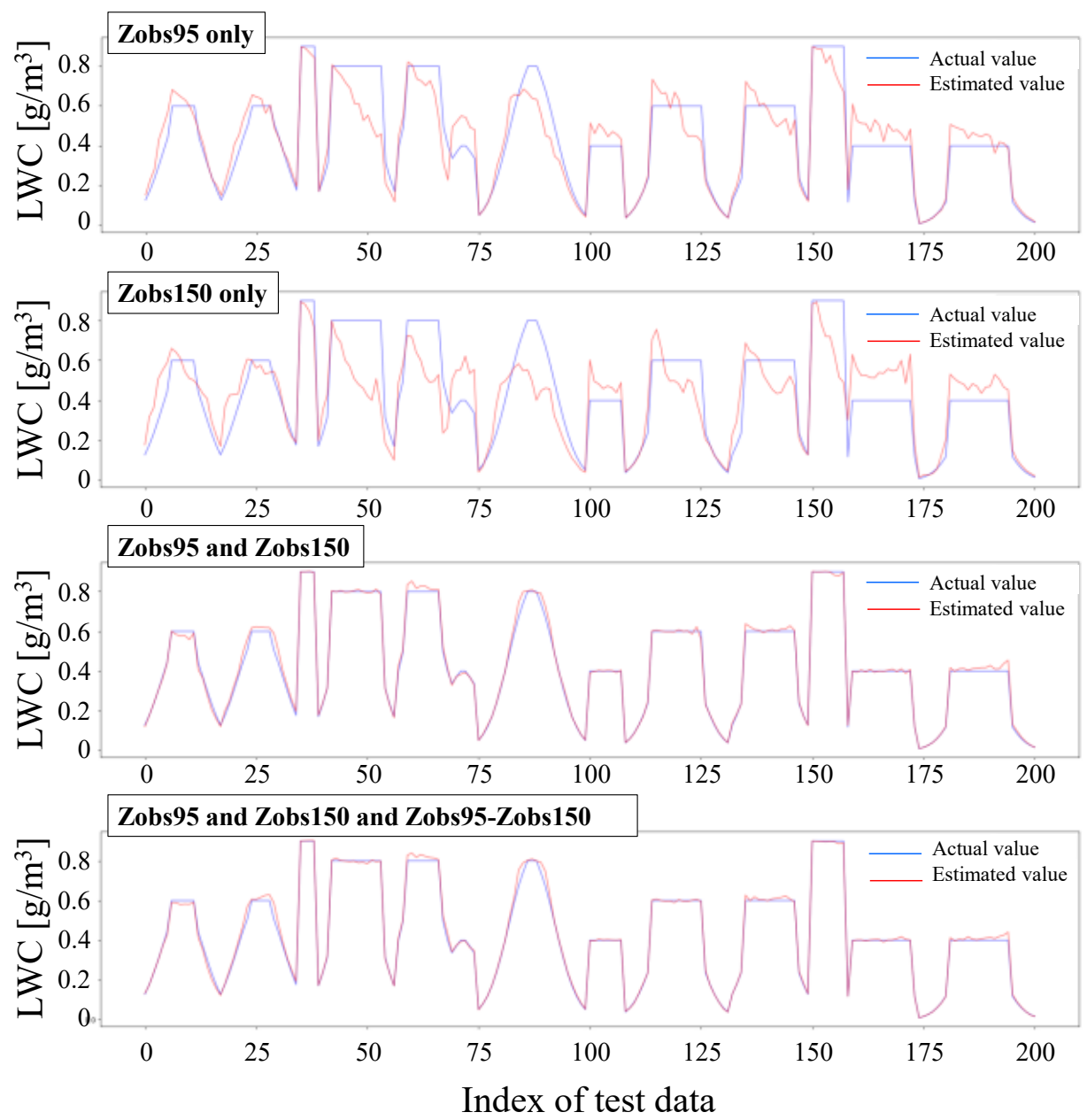

Fig. 9. Estimation Results in Fold 10 (LWC)

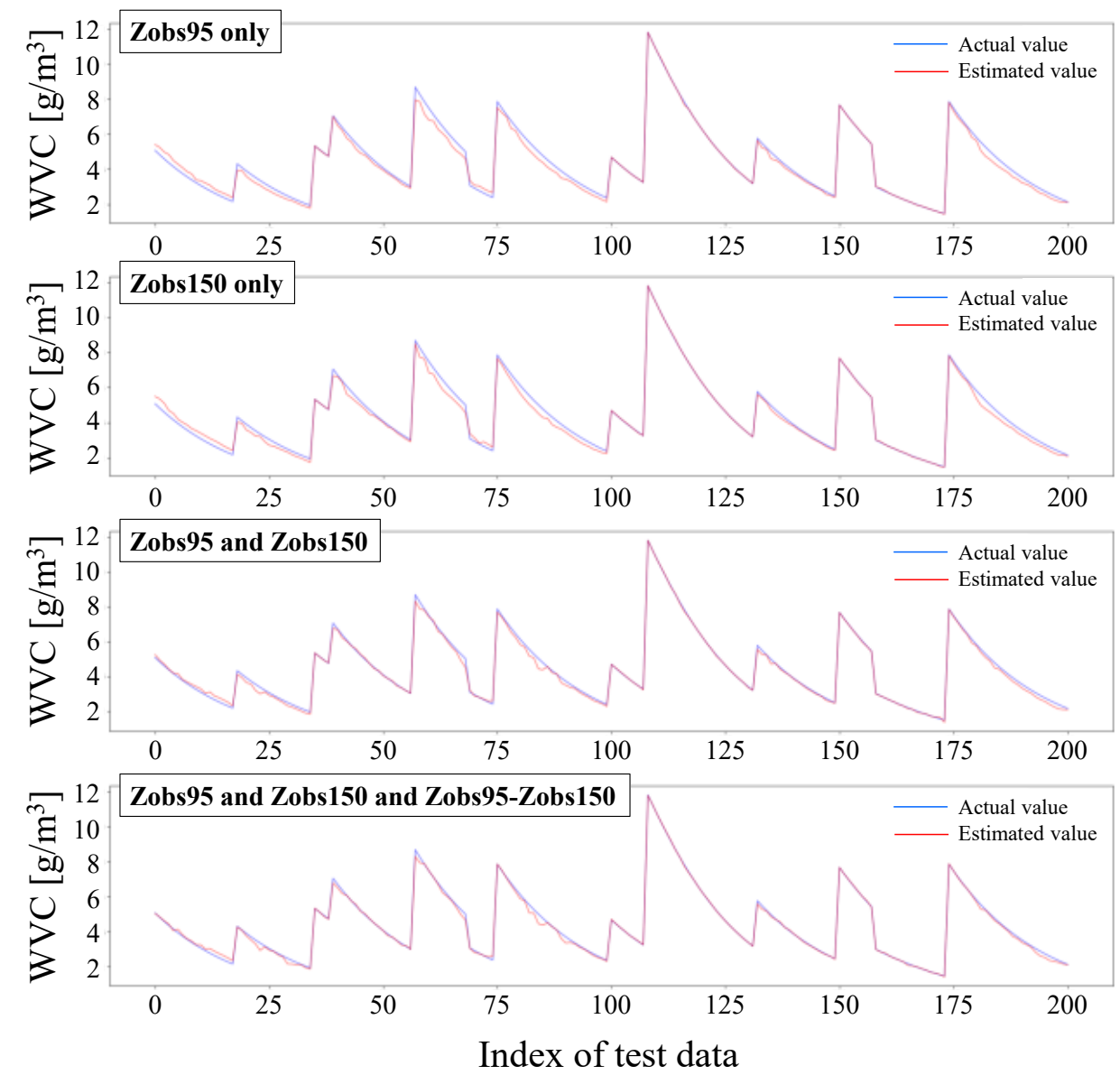

Fig. 10. Estimation Results in Fold 10 (WVC)

## IV. Conclusion

We propose a terahertz dual-frequency cloud radar system using 150 GHz and 95 GHz bands. And from the model constructed by the ITU-R radio propagation model [5, 6, 8], analytically calculated dataset including the true and observed radar reflectivities, LWC and WVC in atmospheric gases are generated. By applying XGBoost to the dataset, 10-fold cross validation is conducted to estimate LWC and WVC. The results show that adding DWR, the difference of radar reflectance between two frequencies, as an explanatory variable improvs the estimation accuracy of LWC and WVC.

In the future, we plan to study the application of publicly available dual-frequency cloud radar data to ML. Specifically, we plan to analyze the measured radar output, to create a data set with radiosonde output, and to evaluate the data with observations obtained from microwave radiometers.

## ACKNOWLEDGMENT

These research results were obtained from the commissioned research (No.06901) by National Institute of Information and Communications Technology (NICT), Japan.